\documentclass[runningheads]{llncs}

\usepackage[T1]{fontenc}
\usepackage[utf8]{inputenc}
\usepackage{xspace}
\usepackage{cite}
\usepackage{amsmath}
\usepackage{proof}
\usepackage{./why3lang}
\usepackage{makecell}
\usepackage{longtable}
\usepackage{multirow}
\usepackage{xspace}
\usepackage[draft]{minted}

\definecolor{thegray}{rgb}{0.9,0.9,0.9}
\definecolor{colorspec}{rgb}{0,0,0.797}
\definecolor{thered}{rgb}{0.797,0,0}
\definecolor{darkgreen}{rgb}{0.797,0,0}
\definecolor{theblue}{rgb}{0,0,0.797}
\definecolor{darkgray}{rgb}{0.8477,0.8477,0.8477}
\definecolor{ocaml-bg}{rgb}{0.9,0.9,0.9}
\definecolor{thegraygray}{rgb}{0.5,0.5,0.5}

\newcommand{\why}{{Why3}\xspace}
\newcommand{\whyml}{{WhyML}\xspace}
\newcommand{\ocaml}{{OCaml}\xspace}

\newcommand{\gospel}{{GOSPEL}\xspace}

\usepackage{graphicx}
\usepackage{hyperref}

\begin{document}
\title{A Deductive Verification Framework For Higher Order Programs}

\author{Tiago Soares --- 52965 \\%[.5em]
Supervisors: Mário Pereira \and António Ravara\\
\email{tl.soares@campus.fct.unl.pt}}
\authorrunning{Tiago Soares --- 52965 \\ Supervisors: Mário Pereira \and António Ravara\\}
\institute{NOVA LINCS \& DI, FCT, Universidade Nova de Lisboa}

\maketitle              % typeset the header of the contribution
\begin{abstract} In this report, we present the preliminary work developed for our research project for the APDC
(\textit{Área Prática de Desenvolvimento Curricular}) course.
The main goal of this project is to develop a framework, on top of the \why tool, for the verification of
effectful higher-order programs. We use defunctionalization as an intermediate transformation from higher-order OCaml implementations into first order ones. The target for our translation is \whyml, the \why's programming language. We believe
defunctionalization can be an interesting route for the automated verification of higher-order programs, since one
can employ off-the-shelf automated program verifiers to prove the correctness of the generated first-order program.
This report also serves to introduce the reader to the subject of deductive program verification and some of the tools
and concepts used to prove higher order effectful programs.

\keywords{Deductive Software Verification  \and Imperative Programs \and Defunctionalization \and Why3 \and OCaml} %Mário: verificar se no restante do relatório dizemos 'stateful' ou 'imperative'
\end{abstract}
\section{Introduction}

Producing error free code for a complex system is, regardless of the programmer’s skills, a herculean task. Customarily, the only weapons we have to combat the plethora of bugs rotting away our code are tests and plenty of caution. But caution can only get you so far and program testing, in Dijkstra's words, "can be a very effective way to show the presence of bugs, but is hopelessly inadequate for showing their absence"\cite{humble}. Therefore, in order to make sure that our code does what we want, we need formal proofs of its correctness.

There is no one definition of correctness, however, considering it depends on how we want to specify our program. We might simply want to make sure that there are no runtime errors involving type safety, arithmetic overflows or unhandled exceptions. We might also want to prove that the program eventually terminates. More generally, we might want to prove some relation to its input and/or the state with its output. For each definition of correctness there are better ways of proving it. For the class of programs we will 
analyse, that is, functional programs, the best alternative %would be
is deductive verification.\cite{filliatre11sttt}

%connector paragraph to demonstrate there are different types of correctness and for each type there are better alternatives. Deductive verification is very suitable for functional correctness

Dispatching proofs for imperative implementations can be challenging,
as they heavily rely on modifying data structures after they have been created, which can be hard to model using first order logic. Therefore, writing in a programming language that stays as close as possible to logical definitions, that is, a functional language, means we will have an easier time identifying the necessary assertions to create an adequate proof. \ocaml is a good candidate for writing effectful, verifiable code, not only because it is primarily a functional language, but also because it allows us to use imperative constructs, such as mutable records. 

To write specifications for code with side effects, we need additional infrastructure on top of first order logic. One of the most popular is Separation Logic~\cite{reynolds2002separation}, which allows us to connect physical memory to logical statements. Although it is very expressive, its intrinsic technicalities prevent it from being widely adopted. In this work, we will be using \gospel (Generic \ocaml SPEcification Language)~\cite{ChargueraudFLP19}, which is built on top of Separation Logic, as a higher level alternative to define changes in the heap and, to prove the specification, we will use \why~\cite{why3-2013}, a program verifier tailored towards automated verification.

But effectful code is only one half of this project; the other being higher-order functions. As previously said, code with side effects can be difficult to map onto logical definitions, but this problem becomes exponentially more complex when these effects originate from first class functions, due to the way programs are proved. In short, we annotate them with pre and postconditions and check that an implementation conforms to the provided specification. This is normally done by employing some technique of verification conditions generation, such as Dijkstra Weakest Precondition Calculus~\cite{wpc}. However, if we don't know the behaviour of a part of our routine, and this unknown part might alter our program's state, then it can be hard to prove if a given program is correct.

Some work has been put into bringing closer automated verification and higher-order programs with effects, such as the tool Who~\cite{KanigFilliatre09wml}. Nonetheless, a framework of that caliber can be complex to implement and maintain. Another option would be Iris ~\cite{jung2018iris}, one of the most powerful tools available in the realm of program verification, but whose power comes at a cost. The logic employed by it is, albeit very rich, extremely complex and almost inaccessible to anyone without years of training in the field.  In addition, there is no way to automate proofs with it, this being the main reason why Iris falls out of the scope of our project.

Most options available for handling this problem present their own hindrances, such as, for example, their complexity, cross-compatibility and their practical usefulness~\cite[Sec. 3.4]{ohearn2018}. We have put forth a small contribution to solve these limitations by creating a \why plugin which makes use of defunctionalization~\cite{danvy2001,reynolds1972} to handle the verification of such programs. Defunctionalization works by translating a higher-order implementation into an equivalent first-order counterpart. We believe this technique can open the door to automatic verification of higher-order code, since we can use an existing first-order verifier to prove the defunctionalized program. 

Previous work studied the combination of defunctionalization and software verification.~\cite{pereira2019}. However, no prototype was developed and all examples were made entirely by hand.
In this project, we lay the groundwork for a fully-fledged verification framework for effectful higher-order programs, by exploring the technique of defunctionalization. To the best of our knowledge, this is the first practical implementation of defunctionalization as a proof technique. Our framework will accept as input a higher-order program, annotated with assertions written
in \gospel, translate it into a first-order equivalent \whyml program so that we can use \why to prove it satisfies the assertions.
 
\section{\why}
\why is a verification framework that allows us to interface with multiple theorem provers and automated solvers using the same surface syntax and underlying logic ~\cite{filliatre2013one,bobot2011why3}. This means that, if one such solver is not able to discharge a verification condition, one can easily switch to another which might be able to complete our proof.

To understand how it works, we will present the defenition of an array list type written in \why's dedicated programming language \whyml: 

\begin{why3}
  type arrayList 'a = {
    mutable arr : array 'a;
    mutable size : int
  }
\end{why3}
This type definition features a field \whyf{arr} of type \whyf{array 'a}, the polymorphic type of arrays defined in \why standard library, and an integer \whyf{size} representing the number of elements in a value of type \whyf{arrayList}. 

Although our type declaration has all we need to build a comprehensive set of functions for our \whyf{arrayList}, from a logical point of view we are missing crucial information. In particular, that the \whyf{size} attribute cannot be greater than the length of the array. To express this, we will add the following type invariant:
\begin{why3}
  invariant { 0 <= size <= arr.length }
\end{why3}
Next, we will write a function that replaces the element at a given index and returns the old element:
\begin{why3}
  let set (al: arrayList 'a) (i : int) (e : 'a) : 'a
\end{why3}
For simplicity’s sake, we will add a \emph{precondition} requiring \whyf{i} to be within the bounds of our list:
\begin{why3}
  requires { 0 <= i < al.size }
\end{why3}
Had we not included such precondition, we would have to include in the body of \whyf{set} an \whyf{if..then..else} expression to avoid accessing an index out of bounds. This a common practice of \emph{defensive programming}.

Next, we will need to specify that the result of our function is the old value at index \whyf{i} and that the new value is \whyf{e}. To do this, we will will introduce two \emph{postconditions}, where we use the \whyf{old} tag to refer to our array list before the function was called and \whyf{result} for our function's return value

\begin{why3}
  ensures  { result = old al.arr[i] }
  ensures  { al.arr[i] = e }
\end{why3}

Lastly, we also need to specify that every other element in the array, as well as its size, is unchanged.
\begin{why3}
    ensures { al.size = old al.size }
    ensures { forall j. 0 <= j < al.size /\ not j = i -> 
                old al.arr[j] = al.arr[j] } 
\end{why3}
Now we simply write the body of the function, trivial as it is:

\begin{why3}
  = let ret = al.arr[i] in 
    al.arr[i] <- e; ret
\end{why3}

The full annotated function can be found in annex \ref{append}. When fed into \why, the above \whyf{set} function generates six verification conditions. These include safety properties (access index in array bounds, which is proved with our invariant) and functional ones (type invariant preservation and postcondition entailment). All of these are automatically verified in less then a second by the Alt-Ergo~\cite{iguer13phd} SMT solver, one of the provers \why allows us to interface with.

\section{Defunctionalization} \label{defun}
Before jumping into our framework's implementation lets first take a gander at a few quick (original) examples to understand how defunctionalization works, how it can be mechanized and how we can apply it to program verification.
\subsection{A simple example}

As previously mentioned, defunctionalization is the practice of turning higher-order programs into equivalent first-order implementations. To illustrate this method, we will convert a higher-order \ocaml program that reverses a list. The code will be written in CPS (Continuation Passing Style)~\cite{danvy2007one}, which, instead of letting the program return a value to the function’s caller, passes the return values as a parameter to a “continuation” function that is given as a first order value. The implementation is as follows:
\begin{minted}{ocaml}
  let rec reverse_aux_cps l k = match l with
    | [] -> k []
    | x::t -> reverse_aux_cps t (fun r -> x  ::  (k r))
    
  let reverse l = reverse_aux_cps l (fun r -> r)
\end{minted}
There are two anonymous functions: the identity function \whyf{(fun r -> r)} and the function which builds the reverse of the list \whyf{(fun r -> x::(k r))}. In order to begin our gentle descent towards a first order version of this program, we introduce a new algebraic data type to represent continuations:
\begin{minted}{ocaml}
  type 'a kont = Kid | Krev of 'a * 'a kont
\end{minted}
The constructors \whyf{Kid} and \whyf{Krev} match, respectively, the identity function and the one that builds the reverse. \whyf{Krev}‘s two extra arguments represent its free variables, namely the appended value and the continuation function.
Next, we need to introduce an apply function which will replace all first order functions calls:
\begin{minted}{ocaml}
  let rec apply kont arg = match kont with
    | Kid -> let r = arg in r
    | Krev(x, k) -> let r = arg in x::(apply r k)
\end{minted}
The defunctionalization process terminates by replacing, in the original code, native function application with
calls to \whyf{apply} and anonymous functions with constructors of the \whyf{kont} data type.
\begin{minted}{ocaml}
  let rec reverse_aux_defun l k = match l with
    | [] -> apply [] k
    | x::t -> reverse_aux_defun t (Krev(x, k))

  let reverse_defun l = reverse_aux_defun l Kid
\end{minted}

The full defunctionalized implementation can be found in annex \ref{revd} and if the reader wishes to understand in more detail how this function would run, we have included an example of the program running in annex \ref{stack}.

\subsection{Proving a defunctionalized program}

With defunctionalization's bases covered, let us introduce how it's used in a proof context. For this, we consider a simpler function, namely one that computes a list's size, also written in CPS. The \whyml implementation is as follows:
\begin{why3}
  let rec length_cps (l : list 'a) (k : int -> 'b) : 'b =
    match l with
    | [] -> k 0
    | _ :: t -> length_cps t (fun l -> k (1 + l))
    end

  let len (l : list 'a) : int =
    length_cps l (fun x -> x)
\end{why3}
Now we annotate our functions with their respective postconditions. The \whyf{len} function's specification is fairly straightforward:
\begin{why3}
  ensures { length l = result }
\end{why3}
where \whyf{length} is a purely logical function, issued from the \why standard library, which can only be used within specifications and shall never be mixed with executable code. 
The CPS function's specification is a bit more intricate. We could, naively, write something like
\begin{why3}
  ensures { k (length l) = result }
\end{why3}
However, this leaves us with the problem of how to interpret the use of functions in our logic, since these may
have side effects (\emph{e.g.}, divergence) which can lead to logical inconsistencies. Therefore, we introduce a layer of
abstraction to allow us to use potentially effectful programs in our assertions: instead of thinking of functions
as a computation of an output from a given input, such as \whyf{f: t1 -> t2}, we will think of them as a pair of
predicates, namely their precondition \whyf{(pre : t1 -> prop)} and postcondition \whyf{(post : t1 -> t2 -> prop)}. This is strongly inspired by the work of Régis-Gianas and Pottier~\cite{regis-gianas-pottier-08}. Under this setting, we would provide with \whyf{length_cps} the following specification:
\begin{why3}
    ensures { post k (length l) result }
\end{why3}
and the continuation defined in \whyf{length_cps} with

\begin{why3}
    ensures { post k (l + 1) result }
\end{why3}
Now we will proceed with our program's defunctionalization as we did before, first by defining a type to represent our first order functions:
\begin{why3}
  type kont = Kid | Klen kont
\end{why3}
The next step, that is, the apply function, is quite similiar to our previous example, which can be found in annex \ref{length}. Finally, since we have no higher order functions, we must define our own post predicate, as follows:
\begin{why3}
  predicate post (k : kont) (length : int) (result : int) = 
    match k with
    | Kid -> result = length
    | Klen k -> post k (length + 1) result
    end
\end{why3}
For the above first-order implementation and specification, all the verification conditions generated by \why are automatically discharged in no time.

As we can observe, the defunctionalized definition of predicate \whyf{post} is a simple matter of collecting the
predicate's uses from the original higher-order implementation. As for the specification of function \whyf{apply}, since this function simulates native application, we provide it with a postcondition stating that applying~\whyf{k} to \whyf{arg} must be directly related to \whyf{result}. As we will show, the translation of the \whyf{post} predicate and the generation of the specification of the \whyf{apply} are indeed amenable to automatic tool translation.

\section{The \gospel to Why3 Pipeline}
    
    In spite of the fact defunctionalization can be a bit of a difficult concept to grasp, once it is understood it becomes clear that we can divide the process into neatly defined steps and that these steps can very easily be mechanized. Although we give a complete overview of how \ocaml code is converted in our translation scheme, found in annex \ref{trans}, we will explain in detail some of the trickier rules and exemplify with some case studies how our converter fares when confronted with decently complex programs. The full implementation for our translation scheme, along with some test cases, can be found in our GitLab \footnote{\href{https://gitlab.com/releaselab/defunctionalization/-/tree/patch-1}{https://gitlab.com/releaselab/defunctionalization/-/tree/patch-1}}

\subsection{Defunctionalization at Work}
 Explaining in detail every single one of our translation scheme's minutia would be a tad redundant, seeing as \whyml and \ocaml are very similar, given they both derive from the ML dialect. Indeed, most of our translations simply amount to lightly re-organizing the \ocaml code to fit \whyml's syntax, picking up the free variables and figuring out the type of the expression. Nevertheless, we will convert a handful of expressions to explain some of the peculiarities of how we approach defunctionalization before we get into the relatively more complex case studies.

 First we will be looking at the translation of the following expression : \whyf{x + y}.
    When converting an \ocaml expression we will return, along with its \whyml translation, its set of free variables and its type. Seeing as this expression belongs to the class $e1\ o\ e2$, \whyf{x} being $e1$, \whyf{+} being o and \whyf{y} being $e2$, we will need $e1$ and $e2$'s translation and free variables. Seeing as  $e1$ and $e2$ are simply two variables, we will translating both directly, meaning e1's translation will be \whyf{x} and  $e2's$ will be \whyf{y} and their set of free variables will be a singleton set each only containing a tuple with the variable's name and its type, which we obtain from a dictionary we refer to in our translation scheme as varType which binds the name of user defined variables to their types.
    
    Finally we need only to combine the two translations with \textit{o}'s, which will simply be~\whyf{+},  
    producing "x + y", and the two sets of free variables, giving us \{(x, int), (y, int)\}. Note however that we do not define this expression's type attribute, seeing as it is only used when translating function applications and we know that an addition can only produce integers, meaning we will never apply the result of this expression as a function.
    
    It is worth noting that in order to use the addition operator in \whyml we need to import the \whyf{int.Int} module. To handle this, we import, along with \whyf{int.Int}, several other libraries so we can naturally use everything in the \ocaml standard library, such as lists and arithmetic operations.
    
    Next, we will bind the previous expression to a lambda and translate it with the  \textit{fun arg* : $\tau$ -> e1} construction
    
    \whyf{fun (x : int) (z : int) : int -> x + y}
    
    Important to note that \whyf{z} is not used and \whyf{y} is not defined in the scope of this expression. First and foremost, we will be adding the function's arguments' type to the aforementioned varType dictionary. After computing  $e1$'s free variables, we must remove $x$ and $z$ from the set. Since $e1$ is our last translated addition (\whyf{x + y}), our new set of free variables is simply $\{(y, int)\}$, since~\whyf{x} is defined in the scope of this expression. Additionally, we will defunctionalize this lambda using the buildApply function, which takes the lambda's free variables, type and arguments' name and returns its defunctionalized alias. Given the previous defunctionalization examples, the reader might be expecting something like this
    
    \begin{why3}
    type kont1 = K int
    
    let apply (k : kont1) (arg1 : int) (arg2 : int)  : int =
      match k with K y -> let x = arg1 in let z = arg2 in x + y
    \end{why3}
    
    Although this would be an accurate translation, it is not fully defunctionalized, due to the fact \ocaml technically does not support functions with more than one argument, which might seem unintuitive seeing as our original lambda has two. When \ocaml parses it, however, it produces a one argument function that returns another one argument function, such as:
    
    \whyf{fun (x : int) -> (fun z : int -> x + y)}
    
    This means we can partially apply arguments and, instead of throwing an error, \ocaml will simply return a lambda where the applied arguments are now free variables. Due to this detail, we defunctionalize our functions in the way \ocaml defines them instead of how the user does. Hence, we will first translate the innermost lambda into \whyf{K1 y x}
    with the generated defunctionalization being similar to:
    
    \begin{why3}
    type kont1 = K1 int int
    
    predicate post1 (k : kont1) (arg :int) (result : int) =
      match k with K1 y x -> let z = arg in true
    
    let apply1 (k : kont1) (arg : int) : int =
      ensures { post1 k arg result }    
      match k with
      | K1 y x -> let z = arg in y + x
    \end{why3}
    
    Seeing as we didn't annotate our original lambda with a postcondition, we fill in the post predicate with the default value \whyf{true}. Before moving on to the outermost lambda, we will bind the type \whyf{int -> int} to the function \whyf{apply1} the the predicate \whyf{post1}.
    
    Now all that is left is to translate the outermost lambda, which is largely identical to the first one, except we now bind \whyf{int -> (int -> int)} to the function \whyf{apply2} and the predicate \whyf{post2} and, seeing as the body of this function will be our previously generated expression, we must append it a specification that describes its return value so we may use this function in future specifications. We then conclude that our post predicate is as follows:
    
    \begin{why3}
    type kont2 = K2 int
    
    predicate post2 (k : kont2) (arg : int) (result : kont1) =
      match k with K2 y -> let x = arg in result = K1 y x
    \end{why3}
 
    In conclusion, we create two \whyf{apply} functions, one that takes the first argument and returns a continuation and another which accepts that continuation along with the final argument returning the result of the original function.
 
    Another detail is that, unlike in our previous examples, we defined two types (\whyf{kont1} and \whyf{kont2}) with one constructor each instead of one type with two constructors. This is due to the fact that we need different \whyf{apply} functions for each different function type since \whyml does not support generalized algebraic data types~\cite{xi2003guarded} as arguments or return values. We cannot put all our constructors in the same definition because that would mean, if we had more than one \whyf{apply} function, their pattern matching wouldn't be exhaustive and \why wouldn't be able to dispatch a proof. 
    
    Finally, we will set this expression's type as \whyf{int -> (int -> int)} and reset the changes made to varType at the start of the translation, that is, remove the binding of \whyf{x} to \whyf{int} and the binding of \whyf{z} to \whyf{int}. Important to note that removing a pairing means that, if there was any other value associated with the key before adding the removed value, that becomes the new value, if not, the binding now becomes undefined.
    
    To conclude, we will now convert the application of a function to a set of arguments: \whyf{(k 3) 4}.
    
    First, we will translate the expression within brackets with the $e1\ e2$ rule. To do so, we will get the translation of $e1$ and $e2$, both of which are directly obtained; their free variables (which are not relevant for this example), and their type, e2's is irrelevant, due to the fact it is a literal value and those are never used as functions, and e1's is obtained in the same manner as before: by checking the current mapping for the variable name \whyf{k}. For the sake of argument, we will assume \whyf{k} is bound to the previous lambda,
    whose type is \whyf{int -> (int -> int)}. 
    
    With both sub expressions now converted, it is time to turn our attention to our original application, which will be converted as follows:
    
    \begin{why3} 
    apply2 k 3
    \end{why3}
    and it's type will be \whyf{int -> int}, seeing as we have already applied \whyf{k}'s first argument.
    
    Next, we do the same thing for the next argument, the only notable difference is that our expression's type is now \whyf{int -> int}, which is bound to the integer 1 and thus we will use the \whyf{apply1} function, giving us:
    
    \begin{why3} 
    apply1 (apply2 k 3) 4
    \end{why3}
    
    As the one might have noticed, the previous expressions were defunctionalized even though they weren't used in a higher order context. This is another quirk of our translation: any and all function declarations and applications, when converted, are also defunctionalized, regardless of whether or not they are used as parameters. The main reason for this is that, when a function is declared we generally have no way of knowing if it is going to be used as a first order value. Despite the fact that we are defunctionalizing more than what we strictly need, translating every function the same way makes the conversion process more uniform and easier to define and implement.
    
    \subsection{Extending \gospel}
    
    Since we have covered the most relevant rules for translating and defunctionalizing \ocaml code, let us take a look at \gospel. We won't go over its syntax in much detail, seeing as its specifications are quite similar to \why's, although there are some interesting divergences. Namely, \gospel annotations are added as comments beginning with the symbol "@" at the tail end of our implementations or, for anonymous functions, between the \whyf{fun} keyword and the function's arguments like so
    
    \begin{minted}{ocaml}
    fun [@gospel {| ensures Q |}] args... -> exp
    \end{minted}
    
    There are some other which are less relevant, for example, having to redefine the name of each argument and the name for the result and the absence of curly brackets, but for the most part we will use similar constructions as we did for \why specifications. Due to this, our \whyml conversions will be almost identical to their \gospel counterparts, with one small difference: the \whyf{post} "predicate". To understand how we have extended \gospel to allow usages of \whyf{post}, let us consider a small example.
    
    \begin{minted}{ocaml}
        (*@ r = f g x
              ensures post (g : int -> int) x r *)
    \end{minted}
    This will be the specification for a function f which takes a function g as a parameter and an integer x. we have omitted the implementation since we will be focusing on how we will deal with the post condition. As previously explained, \whyf{post} is used to avoid directly using non-pure functions (that is, with side-effects) in our specifications. Because \whyf{post} has to work for any function applied to however many arguments of any types, it cannot be defined through \whyml. Instead, as previously shown, we create several \whyf{posts}, one for each \whyf{apply} function and whenever we encounter a \whyf{post} in our specifications, we will replace its application with the respective defunctionalized \whyf{post}. That is why we put predicate in quotation marks; \whyf{post} is not really a predicate nor a function, but more of a meta-predicate: a predicate which, depending on the situation, can be any predicate you need. The only restriction for using it is that the first argument, that is, the function, must have its type explicitly declared so we know which \whyf{post} to use.
    
    Now that we understand the keyword, let us look at how we convert it. For one argument functions it is almost direct. For the aforementioned specification, we would find the \whyf{post} predicate for functions of type \whyf{int -> int}, and replace the call to \whyf{post} with \whyf{post2} from our previous translation: 
    
    \begin{why3}
    ensures { post2 g x r }
    \end{why3}
    
    It gets interesting, however, when we apply \whyf{post} to a function that takes more than one argument. Let us imagine that the \whyf{g} function was of type \whyf{int -> (int -> int)} and was applied to an extra argument y, such as, \whyf{post g x y r}. The translation of such \whyf{post} would be: 
    
    \begin{why3}
    ensures { forall var0 : kont1. post1 g x var0 -> post2 var0 y r }
    \end{why3}
    
    In simpler terms, if we assume the return value for \whyf{g} when applied to \whyf{x} is \whyf{var0}, whose type would be \whyf{kont1}, then when \whyf{var0} is applied to \whyf{y}, it will return \whyf{r}.
    
\section{Case studies} \label{case}
With the most important rules detailed and understood, we will examine how our tool handles relatively complex programs. The following case studies, unlike the ones presented in Section \ref{defun}, are not original and were presented by one of the project's supervisors in their precursor paper on defunctionalized proofs\cite{pereira2019}.

\subsection{Height of a tree}

Firstly, we will look at at a function that computes the height of a binary tree with the following signature

\begin{minted}{ocaml}
let height_tree_cps (t: int tree): int = ...
(*@ r = height_tree t
      ensures (height t) = r *)
\end{minted}

An interesting peculiarity of this program is that it only accepts a tree of integers, even though, intuitively, a height function should work for any type of tree. However, we won't be dealing with any kind of polymorphism, for that would add certain complications when building our constructors \cite{pottierGADTS}. Another thing worth mentioning is that \ocaml does not have a binary tree type, although \whyml does. This means that, even though the \ocaml program we will present technically does not compile on its own unless the programmers defines a tree exactly as it's done in \whyml, it will result in a valid program after the conversion if we import the \whyf{tree.Tree} module. Even so, one could wonder why we simply don't define the tree type in our code instead of going in this roundabout way. The reason is that we are going to need a logical function from \why standard library, namely the height function, in our specification, which we can only apply in conjunction with \whyml's tree type. This solution, flawed as it may be, will be sufficient to prove our program.

First let us define the recursive auxiliary function that performs the calculation

\begin{minted}{ocaml}
let rec height_tree (t: int tree) (k: int -> int) : int
  = match t with
    | Empty -> k 0
    | Node((lt : int tree), _, (rt : int tree)) ->
        height_tree lt (fun [@gospel 
        {| ensures post (k : (integer -> integer)) 
            (1 + max hl (height rt)) result |}]
          (hl : int) : int -> 
        height_tree rt (fun [@gospel 
        {| ensures post (k : (integer -> integer))
            (1 + max hl hr) result |}] (hr : int) : int -> k (1 + max hl hr)))
(*@ r = height_tree t k
      ensures post (k : (integer -> integer)) (height t) r*)
\end{minted}

This one is a bit trickier than our previous examples, due to the nested continuations. In short, we create two continuations, one for each tree branch, which will calculate their heights. Finally, they will compare the two and return the highest value plus 1. Then, all our main function has to do is run the recursive auxiliary function with the identity function.

\begin{minted}{ocaml}
let height_tree_cps (t: int tree): int =
  height_tree t 
    (fun [@gospel {|ensures result = x|}] (x : int) : int -> x)
(*@ r = height_tree t
      ensures (height t) = r*)
\end{minted}

The fully annotated program and its conversion can be found in our GitLab and can be proved using the CVC4 prover, which dispatches the proof in less than a second.

\subsection{\textit{Small-step} interpreter}
The final example we will present is an interpreter for a small language that consists entirely of integer literals and subtractions thereof:

\begin{minted}{ocaml}
            type exp = Const of integer | Sub of exp * exp
\end{minted}

To define how expressions are reduced, we will present two rules, first we assert that a $Sub$ node with two $Const$ can be reduced to a single $Const$:
\begin{center}
$Sub (Const\ v1) (Const\ v2) \xrightarrow{e} Const (v1 - v2)$
\end{center}
Additionally, we will also need an inference rule that allows us to properly parse larger expressions:

\begin{align*}
\infer{C[e] \xrightarrow{e} C[e']}{e \xrightarrow{e} e'}
\end{align*}

C representing the following reduction context:
\[
  \begin{array}{crl}
    C & ::= & \square \\
      & |   & C[\mathtt{Sub} \; \square \; \mathtt{e}] \\
      & |   & C[\mathtt{Sub} \; (\mathtt{Const} \: \mathtt{v}) \; \square]
  \end{array}
\]

Simply put, $C[e]$ represents a context where the expression $e$ has completely replaced the symbol $\square$, meaning our inference rule reads: if an expression $e$ can be reduced to a simpler $e'$, then any context $C[e]$ can be reduced to $C[e']$

With our interpreter clearly described let us move on to how we represent our contexts. Although the most natural translation would be of an algebraic data type with three constructors, one for each of our rules, we will represent them as first order functions that take an expression and return an expression, leading to much cleaner and compact code.

Next, we will implement a function that quite directly translates our inference rule by taking a context $C$ and an expression $e$ which returns a context and a new expression making sure that $C[e] = C'[e']$, $C'$ and $e'$ being the returned context and expression respectively. 

\begin{minted}{ocaml}
let rec decompose_term (e: exp) (c : exp -> exp) : 
  ((exp -> exp) * exp) 
= match e with
  | Sub (Const (v1 : int), Const (v2 : int)) -> (c, e)
  | Sub (Const (v : int), (e : exp)) -> 
      decompose_term e (fun [@gospel 
        {| ensures post (c : (exp -> exp)) (Sub (Const v) x) result |}] 
      (x : exp) : exp -> c (Sub(Const v, x)))

| Sub ((e1 : exp), (e2 : exp)) -> decompose_term e1 
    (fun [@gospel {| ensures post (c : (exp -> exp)) (Sub x e2) result |}]
    (x : exp) : exp -> c (Sub(x, e2))) 

(*@ c_res, e_res = decompose_term e c
      requires not (is_value e)
      ensures  is_redex e_res && 
        forall res. post (c : (exp -> exp)) e res -> 
          post (c_res : (exp -> exp)) e_res res*)
\end{minted}

The first thing one might notice is the fact we only match \whyf{e} with \whyf{Subs} and no \whyf{Const}. When pattern matching is non-exhaustive, \whyml will add an extra branch of the following type \whyf{|_ -> absurd} and attempt to prove that it is never reached. In this case, it will meet this obligation knowing that \whyf{e} is not a value, in other words, not a constant. This knowledge comes from the \whyf{is_value} predicate, which we do not define above. We haven't added its implementation, nor the others we use, seeing as their quite simple. Moreover, functions with preconditions are handled a bit differently: any function presenting these will be converted directly instead of defunctionalized, seeing as there are no interesting case studies of stateless continuations with preconditions. Therefore we will assume that any function with these will never be used as a first order value and therefore has no need in being defucntionalized, thereby simplifying our proof obligations.

Now that we have a function that satisfies $C'[e'] = C[e]$ we need one for the condition $C'[e'] = e$. For this purpose, all we need is to apply our original function with a context so that $C[e] = e$, in other words, the identity function.

\begin{minted}{ocaml}
let decompose (e: exp) : (exp -> exp) * exp =
  decompose_term e 
    (fun [@gospel {| ensures result = x |}] (x : exp) : exp -> x)
(*@ c_res, e_res = decompose e
      requires not (is_value e)
      ensures is_redex e_res && 
              post (c_res : (exp -> exp)) e_res e *)
\end{minted}

With decomposition out of the way, all that's left is to properly evaluate our expressions. To do this, we will define a function which takes an expression and keeps decomposing and performing head reductions, that is, applying the first rule we presented that reduces a $Sub$ into a $Const$, until it reaches a constant:

\begin{minted}{ocaml}
let red (e : exp) : int =
  match e with
  | Const (v : int) -> v
  | _ -> match decompose e with
         ((c : exp -> exp), (r : exp)) -> 
           let r : exp = head_reduction r in red (c r)
(*@ r = red e
      ensures r = eval e *)
\end{minted}

When converted, this program, whose translation also available in our GitLab, is once again successfully proved, this time by the Alt-ergo prover, in less then a second. However to reach it, we need to supply it with the following lemma, which says that for any two expressions, if they evaluate to the same value and are applied to the same context, their evaluations will stay equal relative to one another.

\begin{why3}
lemma post_eval: 
  forall c: exp -> exp, arg1: exp, arg2: exp, r1: exp, r2 : exp.
    eval arg1 = eval arg2 -> post0 c arg1 r1 -> post0 c arg2 r2 ->
    eval r1 = eval r2
\end{why3}

Since this lemma is much broader than our target function's post condition, there could be some value in making it part of \gospel's (and, naturally, \why's) standard library, along with other general axioms regarding higher order functions, so that the programmer has a fairly decent safety net when their provers are unable to reach the required proof obligations. Of course we couldn't put it as is, given that it depends on the \whyf{eval} function, which is specific to our case study. Nonetheless, we could turn it into a kind of meta-lemma where we take a logical function of type \whyf{'a -> 'a} as an argument. To use such a "meta"-lemma to prove our transitive property would be just a matter of instantiating it with the proper 'eval'-like function.

Due to the fact that \gospel does not yet support lemmas, it will have to be placed directly in our translation. An interesting limitation for this lemma is that it could cease to work if we introduced more first order functions of type \whyf{exp -> exp} that are divorced from the contexts we have described. To handle this, our tool would have to know which specific continuations the user is referring to and create a lemma tailored to them. A hypothetical translation that solves this problem can be found in annex \ref{defunc_lemma}.

\section{Limitations} \label{lim}
Due to time constraints, we could only hope to produce a rough prototype that could be used less as a fully fledged tool and more of a proof of concept. Naturally, what we created presents itself with the following non-exhaustive list of restrictions, along with those previously mentioned in Section \ref{case}:

\begin{itemize}
    \item To prove an \ocaml program, we first needed a way to convert the source code into a more digestible format, such as an abstract syntax tree. Two choices emerged; the Typedtree module and the Parsetree module, both from \ocaml's compiler-libs library; the main difference between the two being that the Typedtree has already inferred the types for all expressions and the Parsetree only has the type information that the user puts in the source code. Seeing as, to build our defunctionalized constructors, we need to know the types of the free variables, our options were to simply use the Typedtree or, alternatively, the Parsetree, with the caveat that the user would have to explicitly add the types for all variables defined in 'let' bindings, match cases and function definitions. We decided on using the later seeing as it would be the easiest option to convert programs, although it does pose a restriction on the type of programs we can convert.
    
    Due to this choice, our program does not actually type-check the original code and leaves this to \whyml. This means that, theoretically speaking, we could have a situation where a badly typed program is converted and the resulting \whyml program is correctly typed, leading to a nonsensical specification. Although we believe this situation is impossible, we have no formal proof that this is the case.
    
    \item In order to make the best of the time given, we decided, instead of translating it in its entirety, to focus our efforts onto a subset of the \ocaml language. Although we are missing some very useful syntactic sugar such as module declarations and optional arguments, the constructions allowed by our converter (\emph{e.g.}, match expressions, \whyf{let..in}, \whyf{if..then..else}, lambda expressions) give us plenty of breathing room to write rich, functional code.
    
    \item Due to the fact our apply functions will be mutually recursive, proving termination, while not impossible \cite{pereira2019}
    ,would be quite difficult, therefore, we only prove partial correctness; in other words, we do not prove that these functions are terminating.

    \item Defunctionalization can only be used knowing \textit{a priori} which functions will be used as first order values. Therefore, if a function takes, for example, a lambda of type \whyf{int -> int}, and no functions of this type are defined, then our converter throws an error.
    
    \item Although we hope to eventually use defunctionalization to prove functions with side effects, our prototype can only be used to prove pure programs.
\end{itemize}

\section{Conclusions and future work}
In this report we have proposed a translator that can convert \gospel annotated \ocaml programs into \whyml in order to prove their specification, possibly involving higher order functions. This represents not only the first work involving the verification of \gospel implementations, but also, as far as we know, the first usage of defunctionalization in the context of function specification.

While our tool covers an interesting subset of \ocaml programs, we have only scratched the surface of what defunctionalization is capable of. We will save for future work patching up some of the limitations presented in Section \ref{lim}, namely dealing with effectful programs and polymorphism. To tackle the former, we anticipate the need to by pass the state of the program as an extra parameter to our posts \cite{pereira2019}; for the latter, a much more profound extension to Why3 is needed, namely  extending \why's type system with generalized algebraic data types \cite{pottierGADTS}.

 \bibliography {ms}
\bibliographystyle {splncs04.bst}

\newpage

\appendix

\section{Annotated set function} \label{append}
\begin{why3}
use array.Array
use int.Int

type arrayList 'a = {
    mutable arr : array 'a;    
    mutable size : int
  }
invariant { 0 <= size <= arr.length }

let set (al : arrayList 'a) (i : int) (e : 'a) : 'a
  requires { 0 <= i < al.size }
  ensures  { al.arr[i] = e }
  ensures  { result = old al.arr[i]}
  ensures  { al.size = old al.size}
  ensures  { forall j. 0 <= j < al.size /\ not j = i -> 
  old al.arr[i] = result.arr[i] } 
= let ret = al.arr[i] in
  al.arr[i] <- e;
  ret

\end{why3}

\section{Defunctionalized reverse function} \label{revd}
\begin{minted}{ocaml}
  type 'a kont = Kid | Krev of 'a * 'a kont

  let rec apply kont arg = match kont with
    | Kid -> let r = arg in r
    | Krev(x, k) -> let r = arg in x::(apply r k)

  let rec reverse_aux_defun l k = match l with
    | [] -> apply [] k
    | x::t -> reverse_aux_defun t (Krev(x, k))

  let reverse_defun l = reverse_aux_defun l Kid
\end{minted}

\newpage

\section{Reverse function stack trace} \label{stack}
\begin{minted}{ocaml}
reverse_defun [1;2;3]
\end{minted}

which then has the following stack trace

\begin{minted}{ocaml}
reverse_aux_defun [1;2;3] Kid

reverse_aux_defun [2;3] Krev(1, Kid)

reverse_aux_defun [3] Krev(2, Krev(1, Kid))

reverse_aux_defun [] Krev(3, Krev(2, Krev(1, Kid)))

apply [] Krev(3, Krev(2, Krev(1, Kid)))

3::(apply [] Krev(2, Krev(1, Kid))

3::2::(apply [] Krev(1, Kid))

3::2::1::(apply [] Kid)

3::2::1::[]
\end{minted}

Given this, we can see that this function returns the list 3::2::1::[], that is, [3;2;1], the reverse of [1;2;3]. Interesting to note that for each call to apply the first argument, namely the list, doesn't change. With this in mind, we could have had the apply function take a void value (called \whyf{unit} in ocaml) as its first argument and had the Kid function simply return the empty list. We chose to do it this way, despite this, to keep the \whyf{reverse_cps} function as close to CPS conventions as possible.

\newpage

\section{Defunctionalized length} \label{length}
\begin{why3}
use int.Int
use list.List
use list.Length

  let rec function apply (k : kont) (arg : int) : int
    ensures { post k arg result } =
    match k with
    | Kid -> let x = arg in x
    | Klen kont -> let l = arg in apply kont (1 + l)
    end

 let rec function length_defun (l : list 'a) (k : kont) : int
   ensures { post k (length l) result } =
   match l with
   | Nil -> apply k 0
   | Cons _ t -> length_defun t (Klen k)
   end

  let length_fin (l : list 'a)
    ensures { result = length l } =
  length_defun l Kid
\end{why3}

\newpage

\section{Translation Scheme} \label{trans}

\begingroup
\setlength{\LTleft}{-20cm plus -1fill}
\setlength{\LTright}{\LTleft}
    \setlength\tabcolsep{5pt}
    \begin{longtable}{| c | c | l |}
    \hline
    \endhead
    \hline
    \endfoot
      \hline
      \thead{Symbol} & \thead{Construction} & \thead{Translation} \\
      \hline
      \multirow{1}{4em}{e} & () &  \makecell[l]{e.trans := "()";}
      \\
      \cline{2-3}
      & s &\makecell[l]{e.trans := s; \\
      e.type := varType[s].top();\\
      e.freeVar := [(s, e.type)];}\\
      \cline{2-3}
      & [] & \makecell[l]{e.trans := "Nil"}\\
      \cline{2-3}
      & n & \makecell[l]{e.trans := n}\\
      \cline{2-3}
      & (s)?(e1 (,e2)*) & \makecell[l]{if is\_constructor() then \\
      \qquad c.trans := format("\%s \%s..., s, e1.trans, ...)\\
      else\\
      \qquad c.trans := format("(\%s, ...), e1.trans...) \\
      e.freeVar := union(c1.vars,...)}\\
      \cline{2-3}
      & e1 o e2 & \makecell[l]{e.trans := format("\%s \%s \%s", e1.trans, o.trans, e2.trans);\\e.freeVar := union(e1.freeVar, e2.freeVar) }\\
      \cline{2-3}
      & e1 :: e2 & \makecell[l]{e.trans := format("Cons (\%s) (\%s)",\\ e1.trans, e2.trans);\\
      e.freeVar = union(e1.freeVar, e2.freeVar);}\\
      \cline{2-3}
      & e1;e2 &\makecell[l]{
      e.trans := format("( \%s) ; (\%s)", e1.trans, e2.trans);\\
      e.freeVar := union(e1.freeVar, e2.freeVar);\\
      e.type := e2.type}\\
      \cline{2-3}
      & var\_def in e2  &
      \makecell[l]{e.trans := format("\%s in \%s", var\_def.trans, e2.trans);\\
      e.freeVar := union(var\_def.freeVar,\\ e2.freeVar.except(var\_def.vname));\\
      e.type := e2.type;\\
      varType[var\_def.vname].pop();"}\\
        \cline{2-3}
      & match e1 with (|c  -> e2)+ & \makecell[l]{
      e.trans := format("match \%s with \\
      |\%s -> \%s... end", e1.trans, c.trans, e2.trans, ...);\\
      e.freeVar := union(e1, e2.except(c.vars),..);\\
      e.type := e2.type;
      }\\
      \cline{2-3}
      & if e1 then e2 else e3 & \makecell[l]{
      e.trans := format("if \%s then \%s else \%s",\\
      e1.trans, e2.trans, e3.trans);\\
      e.freeVar := union(e1.freeVar, e2.freeVar, e3.freeVar);\\
      e.type := e2.type
      }\\
      \cline{2-3}
      & fun arg+ : $\tau$ -> e1 & \makecell[l]{
      for all arg do varType[arg.name].push(arg.type);\\
      e.freeVar := e1.freeVar.except(arg1,...);\\
      let name = buildApply(e.freeVar, e1.trans, arg1, ...,\\ $\tau$.trans) in \\
      e.trans := format(\%s(\%s...,), name, e.freeVar[0]...);\\
      e.type := ($\tau$, arg1, ...);\\
      for all arg do varType[arg.name].pop();
      }\\
      \cline{2-3}
      &e1 e2 & \makecell[l]{e.trans := format("\%s \%s \%s", getApply(e1.type), e1.trans, e2.trans\\
      e.freeVar := union(e1.freeVar, e2.freeVar)\\
      e.type := consumeArg(e1.type)}\\
        \hline
        \multirow{5}{4em}{t} 
        &type ('s1 (,'s)*)? s2 = (| s3 (of s4(* s)*)?)+
        &\makecell[l]{t.trans := fomat("type \%s \%s... = |"\%s \%s... =\\ {\%s...}", s2, s1,..., arg.trans)}\\
        \cline{2-3}
        &type ('s1 (,'s)*)? s2 = {arg(;arg)*}&
        \makecell[l]{t.trans := fomat("type \%s \%s... = |"\%s \%s... =\\ {\%s...}", s2, s1,..., arg.trans)}\\
        \cline{2-3}
        &type ('s1 (,'s)*)? s2 = $\tau$&
        \makecell[l]{t.trans := format("type \%s \%s... =\\ \%s, s2, s1,..., $\tau$.trans)}\\
        \hline
        \multirow{1}{4em}{c}
        & n & \makecell[l]{c.trans := n}\\
        \cline{2-3}
        & (s)?(c1 (,c2)*) & 
        \makecell[l]{if is\_constructor() then \\
     \qquad c.trans := format("\%s \%s..., s, c1.trans, ...)\\
else\\
     \qquad c.trans := format("(\%s, ...), c1.trans...) \\
     c.vars := union(c1.vars, c2.vars, ...)}\\
     \cline{2-3}
     &arg&\makecell[l]{l.trans := arg.trans;\\ l.vars = \{(arg.name)\}}\\
     \cline{2-3}
     &[ ]& \makecell[l]{c.trans := "Nil"}\\
     \cline{2-3}
     & c1::c2 & \makecell[l]{l.trans := format("Cons \%s \%s", c1.trans, c2.trans);\\ c1.vars := union(c1.vars, c2.vars)}\\
     \hline
     \multirow{1}{4em}{o}
     & +(.)? & \makecell[l]{o.trans := "+"}\\
     \cline{2-3}
     &-(.)?&\makecell[l]{o.trans := "-"}\\
     \cline{2-3}
     &*(.)?&\makecell[l]{o.trans := "*"}\\
     \cline{2-3}
     &/(.)?&\makecell[l]{o.trans := "/"}\\
     \cline{2-3}
     & \&\& &\makecell[l]{o.trans := "\&\&"}\\
     \cline{2-3}
     &||&\makecell[l]{o.trans := "||"}\\
     \cline{2-3}
     &=&\makecell[l]{o.trans := "="}\\
     \hline
     \multirow{1}{4em}{$\tau$}
     & s & \makecell[l]{$\tau$.trans := s}\\
     \cline{2-3}
     & s (-> $\tau$)+ & $\tau$.trans := \makecell[l]{String.format("\%s -> \%s, s, $\tau$.trans)}\\
     \cline{2-3}
     & () & \makecell[l]{$\tau$.trans :="()"}\\
     \hline
     \multirow{1}{4em}{arg} 
     & () & \makecell[l]{arg.trans := "()"}\\
     \cline{2-3}
     & (s : $\tau$) & \makecell[l]{arg.trans := format("s : \%s", $\tau$.trans);\\
     arg.type := $\tau$.trans;\\
     arg.var := s}\\
     \hline
     \multirow{1}{4em}{var\_def}
     & let (rec)? s arg* : $\tau$ = e1 &
     \makecell[l]{
     varDef.vname := s;
     if isRec() then \\
     \qquad varType[s].push(($\tau$, arg1,...));\\
if isFun() then\\
    \qquad buildApply(s, e1.freeVar, arg1, ..., $\tau$.trans);\\
var\_def.trans := format("let \%s = \%s in ", s, e1.trans);\\
if not isRec() then\\
    \qquad varType[s].push((arg1.type,..., $\tau$));}\\
    \hline
    \multirow{1}{4em}{top\_level} 
    & t;; & \makecell[l]{top\_level.trans := t.trans}\\
    \cline{2-3}
    & var\_def;; & \makecell[l]{top\_level.trans := var\_def.trans}\\
    \cline{2-3}
    & e;; & top\_level.trans := \makecell[l]{format("let \%s = \%s",  VAR\_NAME, e.trans)}\\
    \end{longtable}
    
    \endgroup
\newpage

\section{Defunctionalized lemma} \label{defunc_lemma}
\begin{why3}
  let rec lemma post_eval (k: kont0) (arg1 arg2 r1 r2: exp)
    requires { eval arg1 = eval arg2 }
    requires { post0 k arg1 r1 && post0 k arg2 r2 }
    ensures  { eval r1 = eval r2 }
    variant  { k }
  = match k with
    | K2 -> ()
    | K1 e c ->
       post_eval c (Sub arg1 e) (Sub arg2 e) r1 r2
    | K0 n c ->
       post_eval c (Sub (Const n) arg1) (Sub (Const n) arg2) r1 r2
    |_ -> true
    end
\end{why3}

\end{document}